\documentclass[12pt]{article}
\usepackage{amssymb,amsmath,epsfig}

\begin{document}
\title{\bf Structure Scalars for Charged Cylindrically Symmetric Relativistic Fluids}
\author{M. Sharif \thanks{msharif.math@pu.edu.pk} and M. Zaeem Ul
Haq Bhatti \thanks{mzaeem.math@gmail.com}\\
Department of Mathematics, University of the Punjab,\\
Quaid-e-Azam Campus, Lahore-54590, Pakistan.}

\date{}

\maketitle
\begin{abstract}
We investigate some structure scalars developed through Riemann
tensor for self-gravitating cylindrically symmetric charged
dissipative anisotropic fluid. We show that these scalars are
directly related to the fundamental properties of the fluid. We
formulate dynamical-transport equation as well as the mass function
by including charge which are then expressed in terms of structure
scalars. The effects of electric charge are investigated in the
structure and evolution of compact objects. Finally, we show that
all possible solutions of the field equations can be written in
terms of these scalars.
\end{abstract}

\section{Introduction}

It is believed that anisotropy plays a vital role for understanding
the gravitation of those objects which have higher densities than
neutron stars. Many phenomena such as the solid core, phase
transition, mixture of two fluids, slow rotation and pion
condensation can generate anisotropy in the star models \cite{1}.
Lemaitre \cite{6} was the first who gave the idea about the
tangential and radial pressures. Since the pioneering work of Bowers
and Liang \cite{7}, there has been extensive literature devoted to
the study of general anisotropic relativistic configuration both
analytically and numerically \cite{8}.

Recent literature indicates interesting consequences of the
inclusion of an electromagnetic field to discuss gravitational
collapse. Bekenstein \cite{50} extended the Oppenheimer-Volkoff
equations of hydrostatic equilibrium \cite{51} from the neutral to
the charged case. Herrera et al. \cite{51a} provided a set of
equations for the physical interpretation of models for collapsing
charged spheres. Nath et al. \cite{52} discussed charged
gravitational collapse and concluded that electromagnetic field
increases the formation of naked singularity. Sharif and his
collaborators \cite{53} investigated the effects of electromagnetic
field on different aspects of spherically/cylindically and plane
symmetric gravitational collapse.

The orthogonal splitting of the Riemann tensor was first considered
by Bell \cite{20}. Herrera and his collaborators \cite{21} followed
this idea to develop a relationship between structure scalars and
the fluid properties. Also, they \cite{25} analyzed the structure
scalars for charged dissipative spherical fluids. Structure scalars,
$X_T,~X_{TF},~Y_T,~Y_{TF}$ have important properties such as $X_T$
is the energy density of the fluid, $X_{TF}$ controls inhomogeneity
in the fluid, $Y_{TF}$ describes the effect of the local anisotropy
of pressure as well as density inhomogeneity of the Tolman mass and
$Y_T$ turns out to be proportional to the Tolman mass density for
systems in equilibrium or quasi-equilibrium. This ultimately
provides a relationship between structure scalars and energy density
inhomogeneity.

In a recent paper, Herrera et al. \cite{26} discussed cylindrically
symmetric relativistic dissipative fluids based on structure
scalars. Here we take the effect of electromagnetic field to study
the structure scalars with the same configuration. The paper is
organized as follows. In the next section, we describe the
Einstein-Maxwell field equations, kinematics and the Weyl tensor.
Section \textbf{3} is devoted for the structure scalars and
conservation laws. In section \textbf{4}, we formulate the dynamical
equations in terms of mass function and couple with transport
equation. We discuss the possible static charged cylindrically
symmetric solutions for the anisotropic fluid in section \textbf{5}.
In the last section, we summarize the results.

\section{Charged Anisotropic Dissipative Fluid Cylinders}

Here we review the basic general equations and some definitions. The
general cylindrically symmetric spacetime is given as
\begin{equation}\label{1}
ds^2=-A^2(t,r)(dt^{2}-dr^{2})+B^2(t,r)dz^2 +C^2(t,r)d\phi^2,
\end{equation}
where $-\infty\leqslant{t}\leqslant\infty,~0\leqslant{r},~-\infty<
{z}<\infty,~0\leqslant{\phi}\leqslant2\pi$. We assume that the
collapsing cylinder is filled with anisotropic and dissipative fluid
for which the energy-momentum tensor is
\begin{equation}\label{2}
T^{(m)}_{\alpha\beta}=(\mu+P_{r})V_\alpha V_\beta+P_r
g_{\alpha\beta}+q_\alpha V_\beta+q_\beta V_\alpha+\Pi_{\alpha\beta},
\end{equation}
where $\Pi_{\alpha\beta}=(P_z-P_r)S_{\alpha}S_\beta+(P_\phi-
P_r)K_{\alpha}K_\beta,~q_\alpha = qL_\alpha,~\mu$ is the energy
density, $P_r,~P_z,~P_\phi$ are the pressure in the radial, $z$ and
$\phi$ directions, respectively, $q_\alpha$ is the radial heat flux
and $V_\alpha$ is the four velocity. Also, $S_\alpha,~K_\alpha$ and
$L_\alpha$ are the unit four-vectors with
\begin{eqnarray}\nonumber
V^{\alpha}V_{\alpha}&=&-1,\quad L^\alpha L_\alpha=S^\alpha
S_\alpha=K^\alpha K_\alpha=1,\\\nonumber V^\alpha
L_\alpha&=&V^\alpha S_\alpha=V^\alpha K_\alpha=S^\alpha K_\alpha=0.
\end{eqnarray}
For comoving coordinate system, we have
\begin{equation*}\label{5}
V_{\alpha}=-A\delta^{0}_{\alpha},\quad
L_{\alpha}=A\delta^{1}_{\alpha},\quad
S_{\alpha}=B\delta^{2}_{\alpha}, \quad
K_{\alpha}=C\delta^{3}_{\alpha}.
\end{equation*}

The energy-momentum tensor for electromagnetic field is
\begin{equation}\label{6}
T^{(em)}_{\alpha\beta}=\frac{1}{4\pi}\left(F^{\gamma}_{\alpha}F_{\beta\gamma}
-\frac{1}{4}F^{\gamma\delta}F_{\gamma\delta}g_{\alpha\beta}\right),
\end{equation}
where $F_{\alpha\beta}=\phi_{\beta,\alpha}-\phi_{\alpha,\beta}$ is
the Maxwell field tensor and $\phi_\alpha$ represents the four
potential. The Maxwell field equations are
\begin{equation}\label{8}
F^{\alpha\beta}_{;\beta}={\mu}_{0}J^{\alpha},\quad
F_{[\alpha\beta;\gamma]}=0,
\end{equation}
where $\mu_0=4\pi$ is the magnetic permeability and $J^\alpha$ is
the four current. In comoving coordinate system, we can assume that
the charge per unit length of the system is at rest, so the magnetic
field will be zero. Thus one can choose the four potential and four
current as
\begin{equation*}\label{9}
\phi_{\alpha}={\phi}{\delta^{0}_{\alpha}},\quad
J^{\alpha}={\zeta}V^{\alpha},
\end{equation*}
here $\phi$ is the scalar potential and $\zeta$ is the charge
density, both are functions of $t$ and $r$. The only non-zero
component of the Maxwell field tensor is
\begin{equation*}\label{10}
F_{01}=-F_{01}=-\phi',
\end{equation*}
where prime is the differentiation with respect to $r$. Using these
values, the Maxwell field equations become
\begin{eqnarray}\label{11}
&&{\phi}''+\phi'\left(\frac{B'}{B}+\frac{C'}{C}-\frac{2A'}{A}\right)=4\pi{\zeta}A^3,\\\label{12}
&&\frac{\partial}{\partial
t}\left(\frac{1}{A^4}\frac{\partial\phi}{{\partial}r}\right)
+\left(\frac{1}{A^4}\frac{\partial\phi}{{\partial}r}\right)\left(\frac{\dot{B}}{B}
+\frac{\dot{C}}{C}+\frac{2\dot{A}}{A}\right)=0,
\end{eqnarray}
where dot means differentiation with respect to $t$.
Integration of Eq.(\ref{11}) yields
\begin{equation*}
\phi'=\frac{2sA^2}{BC},
\end{equation*}
where
\begin{equation}\label{13}
s(r)=2\pi\int^r_{0}{\zeta}{ABC}dr,
\end{equation}
is the total amount of charge per unit length of the cylinder found
through the conservation equation, $J^\mu_{;\mu}=0.$ Also, $\phi'$
identically satisfies Eq.(\ref{12}). The Einstein-Maxwell field
equations yield
\begin{eqnarray}\nonumber
{\kappa}A^2\left(\mu+\frac{s^2}{2{\pi}B^2C^2}\right)
&=&\left(\frac{\dot{B}}{B}+\frac{\dot{C}}{C}\right)\frac{\dot{A}}{A}
+\frac{\dot{B}}{B}\frac{\dot{C}}{C}-\frac{B''}{B}-\frac{C''}{C}\\\label{17}
&+&\frac{A'}{A}\left(\frac{B'}{B}+\frac{C'}{C}\right)-\frac{B'}{B}\frac{C'}{C},\\\nonumber
-{\kappa}qA^2&=&-\frac{\dot{B'}}{B}-\frac{\dot{C'}}{C}
+\frac{\dot{A}}{A}\left(\frac{B'}{B}+\frac{C'}{C}\right)\\\label{18}
&+&\frac{A'}{A}\left(\frac{\dot{B}}{B}\right.+\left.\frac{\dot{C}}{C}\right),\\\nonumber
{\kappa}A^2\left(P_r-\frac{s^2}{2{\pi}B^2C^2}\right)
&=&-\frac{\ddot{B}}{B}-\frac{\ddot{C}}{C}+\frac{\dot{A}}{A}\left(\frac{\dot{B}}{B}
+\frac{\dot{C}}{C}\right)-\frac{\dot{B}}{B}\frac{\dot{C}}{C}\\\label{19}
&+&\frac{A'}{A}\left(\frac{B'}{B}+\frac{C'}{C}\right)+\frac{B'}{B}\frac{C'}{C},
\end{eqnarray}
\begin{eqnarray}\nonumber
{\kappa}B^2\left(P_z+\frac{s^2}{2{\pi}B^2C^2}\right)
&=&\left(\frac{B}{A}\right)^2\left[-\frac{\ddot{A}}{A}-\frac{\ddot{C}}{C}
+\left(\frac{\dot{A}}{A}\right)^2+\frac{A''}{A}+\frac{C''}{C}\right.\\\label{20}
&-&\left.\left(\frac{A'}{A}\right)^2\right],\\\nonumber
{\kappa}C^2\left(P_\phi+\frac{s^2}{2{\pi}B^2C^2}\right)
&=&\left(\frac{C}{A}\right)^2\left[-\frac{\ddot{A}}{A}-\frac{\ddot{B}}{B}
+\left(\frac{\dot{A}}{A}\right)^2+\frac{A''}{A}+\frac{B''}{B}\right.\\\label{21}
&-&\left.\left(\frac{A'}{A}\right)^2\right].\\\nonumber
\end{eqnarray}

There are four kinematical variables for the description of fluid,
i.e., expansion, acceleration, shear and rotation. Since we are
using irrotational fluid, so the rotation will be zero. Rest are
defined as
\begin{eqnarray}\nonumber
\Theta=V^{\alpha}_{;\alpha},\quad
a_\alpha=V_{\alpha;\beta}V^\beta,\quad
\sigma_{\alpha\beta}=V_{(\alpha;\beta)}+a_{(\alpha}V_{\beta)}
-\frac{1}{3}\Theta h_{\alpha\beta},
\end{eqnarray}
where $h_{\alpha\beta}=g_{\alpha\beta}+V_{\alpha}V_\beta$. Using
Eq.(\ref{1}), these quantities turn out to be
\begin{equation*}
\Theta=\frac{1}{A}\left(\frac{\dot{A}}{A}+\frac{\dot{B}}{B}
+\frac{\dot{C}}{C}\right),\quad a_1=\frac{A'}{A},\quad
a^2=a^{\alpha}a_\alpha=\left(\frac{A'}{A^2}\right)^2,\quad
a_\alpha=aL_\alpha.
\end{equation*}
The shear tensor can also be expressed as
\begin{eqnarray}\nonumber
\sigma_{\alpha\beta}&=&\sigma_s\left(S_{\alpha}S_\beta-\frac{1}{3}h_{\alpha\beta}\right)
+\sigma_k\left(K_\alpha
K_\beta-\frac{1}{3}h_{\alpha\beta}\right),\\\label{30}
\sigma^{\alpha\beta}\sigma_{\alpha\beta}
&=&\frac{2}{3}(\sigma^2_s-\sigma_s\sigma_k+\sigma^2_k),
\end{eqnarray}
where
\begin{equation*}
\sigma_s=\frac{1}{A}\left(\frac{\dot{B}}{B}
-\frac{\dot{A}}{A}\right),\quad
\sigma_k=\frac{1}{A}\left(\frac{\dot{C}}{C}
-\frac{\dot{A}}{A}\right).
\end{equation*}

The Weyl tensor is defined as
\begin{equation*}\nonumber
C^\rho_{\alpha\beta\mu}=R^\rho_{\alpha\beta\mu}-\frac{1}{2}R^\rho_{\beta}g_{\alpha\mu}
+\frac{1}{2}R_{\alpha\beta}{\delta^{\rho}_{\mu}}-\frac{1}{2}
R_{\alpha\mu}{\delta^{\rho}_{\beta}}+\frac{1}{2}R^\rho_\mu
g_{\alpha\beta}+\frac{1}{6}({\delta^{\rho}_{\beta}}g_{\alpha\mu}
-g_{\alpha\beta}{\delta^{\rho}_{\mu}})
\end{equation*}
which may be decomposed in its electric and magnetic parts as
\begin{equation}\label{32}
E_{\alpha\beta}=C_{\alpha\nu\beta\delta} V^{\nu} V^{\delta},\quad
H_{\alpha\beta}=\frac{1}{2}\eta_{\alpha\nu\epsilon\rho}
C_{\beta\delta}^{\epsilon\rho}V^{\nu}V^{\delta},
\end{equation}
respectively, where $\eta_{\alpha\nu\epsilon\rho}$ is the
Levi-Civita tensor. These can also be written as
\begin{equation}\label{34}
E_{\alpha\beta}=E_s(S_{\alpha}S_\beta-\frac{1}{3}h_{\alpha\beta})
+E_k(K_{\alpha}K_\beta-\frac{1}{3}h_{\alpha\beta}),~
H_{\alpha\beta}=H(S_{\alpha}K_\beta+S_{\beta}K_\alpha),
\end{equation}
where
\begin{equation}\label{36}
E_s=\frac{1}{A^2B^2}C_{0202}-\frac{1}{A^2}C_{0101},~
E_k=\frac{1}{A^2C^2}C_{0303}-\frac{1}{A^2}C_{0101},~
H=-\frac{C_{0313}}{A^2C^2}.
\end{equation}
The components of the Weyl tensor $C_{0202},~C_{0101}$,
$C_{0303},~C_{0313}$ are given in a recent paper \cite{26}.

\section{Structure Scalars}

In this section, we formulate structure scalars for the charged
fluid from the orthogonal splitting of the Riemann tensor \cite{26}.
For this purpose, the following tensors are defined
\begin{equation*}\label{38}
Y_{\alpha\beta}=R_{\alpha\nu\beta\delta}V^{\nu}V^{\delta},\quad
X_{\alpha\beta}=^{*}R^{*}_{\alpha\gamma\beta\delta}V^{\gamma}V^{\delta}=
\frac{1}{2}\eta^{\epsilon\rho}_{\alpha\gamma}R^{*}_{\epsilon\rho\beta\delta}V^{\gamma}V^{\delta},
\end{equation*}
where $R^{*}_{\alpha\beta\gamma\delta}=
\frac{1}{2}\eta_{\varepsilon\rho\gamma\delta}R^{\epsilon\rho}_{\alpha\beta}$.
These can be expressed in trace and trace free parts as
\begin{eqnarray}\nonumber
Y_{\alpha\beta}&=&\frac{1}{3}Y_{T}h_{\alpha\beta}+Y_s( S_\alpha
S_\beta-\frac{1}{3}h_{\alpha\beta})+Y_k(K_{\alpha}K_\beta-\frac{1}{3}h_{\alpha\beta}),\\\label{42}
X_{\alpha\beta}&=&\frac{1}{3}X_{T}h_{\alpha\beta}+X_s(S_{\alpha}S_\beta
-\frac{1}{3}h_{\alpha\beta})+X_k(K_{\alpha}K_\beta-\frac{1}{3}h_{\alpha\beta}),
\end{eqnarray}
Using the field equation (\ref{17})-(\ref{21}) with (\ref{36}), we
obtain $Y_{\alpha\beta}$ and $X_{\alpha\beta}$ in terms of physical
variables
\begin{eqnarray}\label{43}
Y_T&=&\frac{\kappa}{2}(\mu+P_z+P_\phi+P_r)+\frac{4s^2}{B^2C^2},\quad
X_T=\kappa\mu+ \frac{4s^2}{B^2C^2},\\\nonumber
Y_s&=&E_s-\frac{\kappa}{2}(P_z-P_r)-\frac{4s^2}{B^2C^2},\quad
Y_k=E_k-\frac{\kappa}{2}(P_\phi-P_r)-\frac{4s^2}{B^2C^2},\\\nonumber
X_s&=&-E_s-\frac{\kappa}{2}(P_z-P_r)-\frac{4s^2}{B^2C^2},\quad
X_k=-E_k-\frac{\kappa}{2}(P_\phi-P_r)-\frac{4s^2}{B^2C^2}.\\\label{44}
\end{eqnarray}

The conservation law, $T^{\alpha\beta}_{;\beta}=0$, yields
\begin{eqnarray}\nonumber
\mu^*+\Theta(\mu+P_r)+q^\alpha_{;\alpha}
+a_{\alpha}q^\alpha+\sigma_{\alpha\beta}\Pi^{\alpha\beta}
+\frac{1}{3}\Theta\Pi^\alpha_\alpha=0,\\\label{50}
h^{\alpha\beta}(P_{r;\beta}+\Pi^\mu_{\beta;\mu}+q^*_\beta)
+(\mu+P_r)a^\alpha+\frac{4}{3}{\Theta}q^\alpha+\sigma^\alpha_{\mu}q^\mu-\frac{ss'}{{\pi}AB^2C^2}=0.
\end{eqnarray}
The last equation can be written in an alternative form as
\begin{eqnarray}\nonumber
P^\dag_r+q^*&-&\frac{1}{A}\left[(P_z-P_r)\frac{B'}{B}\right.
+(P_\phi-P_r)\left.\frac{C'}{C}\right]+(\mu+P_r)a\\\label{51}
&-&\frac{1}{3}(\sigma_s+\sigma_k-4\Theta)q-\frac{ss'}{{\pi}AB^2C^2}=0,
\end{eqnarray}
where $f^*=f_{,\alpha}V^\alpha,\quad f^\dag=f_{,\alpha}L^\alpha$.
There are two important differential equations relating the Weyl
tensor to different physical variables. Herrera \cite{26} found
these relations for cylindrically symmetric spacetime which are
generalized to charge distribution as
\begin{eqnarray}\nonumber
&&-(Y_s+Y_k-X_s-X_k)^\dag-3(Y_s-X_s)\frac{B'}{AB}-3(Y_k-X_k)\frac{C'}{AC}\\\nonumber
&&-6H(\sigma_s-\sigma_k)=\kappa(2\mu+P_r+P_z+P_\phi)^\dag+3\kappa(\mu+P_r)a\\\label{53}
&&+2{\kappa}q(\Theta-\sigma_s-\sigma_k)+3{\kappa}q^*+\frac{3{\kappa}s}{\pi
AB^2C^2} \left(s'-s\frac{B'}{B}-s\frac{C'}{C}\right),\\\nonumber
&&(2Y_s-Y_k-2X_s+X_k)^\dag+3(Y_s-X_s)\frac{B'}{AB}+3a(Y_s-Y_k-X_s\\\nonumber
&&+X_k)+6H(\Theta-\sigma_k)+6H^*=-\kappa(\mu-P_r-P_z+2P_\phi)^\dag\\\label{54}
&&-3\kappa(P_\phi-P_r)\frac{C'}{AC}+\kappa
q(\Theta-\sigma_s+2\sigma_k)+\frac{3{\kappa}s}
{{\pi}AB^2C^2}\left(s\frac{B'}{B}-s'\right).
\end{eqnarray}

\section{Mass Function and Dynamical-Transport Equation}

Here, we develop equations that govern the dynamics of non-adiabatic
cylindrically symmetric collapsing process. For this purpose, we
define the velocity $U=\frac{\dot{C}}{A}=C^*$. Using Eq.(\ref{19}),
we get
\begin{equation*}\label{55}
U^*=a\frac{C'}{A}-\kappa P_rC-\frac{C}{A^2}\left(\frac{\ddot{B}}{B}
-\frac{\dot{A}\dot{B}}{AB}+\frac{\dot{B}\dot{C}}{BC}-\frac{B'C'}{BC}
-\frac{A'B'}{AB}\right)+\frac{\kappa s^2}{2{\pi}B^2C},
\end{equation*}
which turns out to be
\begin{equation*}\label{56}
U^*=a\frac{C'}{A}-{\kappa}P_rC+\frac{C}{B^2}\left(\frac{R_{0202}}{A^2}
-\frac{R_{2323}}{C^2}\right)+\frac{\kappa s^2}{2{\pi}B^2C}.
\end{equation*}
Solving it for $a$ and substituting into Eq.(\ref{51}),
we obtain
\begin{eqnarray}\nonumber
(\mu+P_r)U^*&=&-(\mu+P_r)\left[{\kappa}P_rC
-\frac{C}{B^2}\left(\frac{R_{0202}}{A^2}-\frac{R_{2323}}{C^2}\right)\right]
-\frac{C'}{A}\left[P^\dag_r\right.\\\nonumber
&-&\left.(P_z-P_r)\frac{B'}{AB}-(P_\phi-P_r)\frac{C'}{AC}\right]
+\frac{C'}{A}\left[-q^*+\frac{1}{3}(\sigma_s\right.\nonumber\\
&+&\left.\sigma_k-4\Theta)q\right]+\left[\frac{ss'C'}{{\pi}B^2C^2}
+(\mu+P_r)\frac{{\kappa}s^2}{2{\pi}B^2C}\right].\label{57}
\end{eqnarray}
This equation yields the effect of different forces on the
collapsing process and have the "Newtonian" form as
\begin{equation*}
Force=Mass~Density~\times~Acceleration.
\end{equation*}
The term on the left is the density (the inertial or passive
gravitational mass density) multiplied by the proper time derivative
of the velocity $U$. The terms on the right represent the force
which is the contribution of four forces: the gravitational force
(the pressure gradient plus the anisotropic contribution), the
contribution from the dissipation and the contribution from the
charge term.

Now we define a mass function $m$ similar to the spherically
symmetric case. For this purpose, we write the term
$\frac{R_{0202}}{A^2}-\frac{R_{2323}}{C^2}$ in the form of structure
scalars as
\begin{equation*}\label{63}
\frac{R_{0202}}{A^2}-\frac{R_{2323}}{C^2}=\frac{B^2}{3}(Y_T-X_T+X_s+2Y_s+X_k-Y_k).
\end{equation*}
Using Eqs.(\ref{43}) and (\ref{44}), it follow that
\begin{equation}\label{64}
\frac{R_{0202}}{A^2}-\frac{R_{2323}}{C^2}=\frac{{\kappa}B^2}{3}\left(-\frac{\mu}{2}
+2P_r-P_z+\frac{P_\phi}{2}\right)+\frac{B^2}{3}(E_s-2
E_k)-\frac{4s^2}{C^2}.
\end{equation}
Inserting this value in the first square brackets on the right side
of Eq.(\ref{57}), we obtain
\begin{eqnarray}\nonumber
\kappa P_rC -\frac{C}{B^2}\left(\frac{R_{0202}}{A^2}
-\frac{R_{2323}}{C^2}\right)=\frac{\kappa
P_rC}{2}&+&\frac{\kappa}{6}(\mu-P_\phi-P_r+2P_z)C\\\label{65}
&-&\frac{C}{3}(E_s-2E_k)+\frac{4s^2}{B^2C}.
\end{eqnarray}

To define a mass function, we have followed the procedure of
\cite{51a,26}. We have compared Eq.(\ref{65}) with the corresponding
equation in the spherically symmetric case \cite{51a}. We have also
assumed that the pressure effect remains the same as in spherically
symmetric case which provides the definition of mass function as a
possible extension of the Misner-Sharp mass function to the
cylindrically symmetric case given by
\begin{equation}\label{66}
m=\frac{C^3\kappa}{6}(\mu-P_\phi-P_r+2P_z)-\frac{C^3}{3}(E_s-2
E_k)+\frac{4s^2C}{B^2}.
\end{equation}
Making use of Eq.(\ref{44}), this can be written as
\begin{equation}\label{67}
\frac{3m}{C^3}=\frac{\kappa}{2}(\mu+P_\phi-2P_r+P_z)-(Y_s-2Y_k)+\frac{16s^2}{B^2C^2},
\end{equation}
or
\begin{equation}\label{68}
\frac{3m}{C^3}=\frac{\kappa}{2}(\mu-3P_\phi+3P_z)-(2X_k-X_s)+\frac{8s^2}{B^2C^2}.
\end{equation}

The mass function in terms of electric charge can be found by using
Eqs.(\ref{44}) and (\ref{51})-(\ref{54}) as follows
\begin{eqnarray}\nonumber
&&(3Y_s-3Y_k+X_s+X_k)^\dag=\kappa\mu^\dag+\kappa
q(\sigma_k-\Theta-2\sigma_s)-\frac{3B'}{AB}(Y_s+X_s)\\\nonumber
&&+\frac{3C'}{AC}(Y_k-X_k)-6H(\Theta
-\sigma_s)-6H^*-3a(Y_s-Y_k-X_s+X_k)\\\label{69}
&&+\frac{3{\kappa}s}{{\pi}AB^2C^2}\left(s'-\frac{sC'}{C}\right).
\end{eqnarray}
Adding Eqs.(\ref{67}) and (\ref{68}), it follows that
\begin{equation*}\label{70}
\frac{6m}{C^3}=(\kappa\mu+3Y_k-3Y_s-X_k-X_s)+\frac{16s^2}{B^2C^2}.
\end{equation*}
Applying the operator $\dag~(f^\dag=f_{,\alpha}L^\alpha)$ on both
sides of the above equation and then substituting in Eq.(\ref{69}),
after some manipulation, we obtain
\begin{eqnarray*}\nonumber
\left(\frac{6m}{C^3}\right)^\dag&=&3(Y_s+X_s)\frac{B'}{A B}
+6H(\Theta-\sigma_s)-3(Y_k-X_k)\\\nonumber
&\times&\frac{C'}{AC}+{\kappa}q(2\sigma_s-\sigma_k+\Theta)+6H^*+3a(Y_s-Y_k\\\label{71}
&-&X_s+X_k)-\frac{{\kappa}s}{{\pi}AB^2C^2}\left(s'-\frac{sC'}{C}-\frac{4sB'}{B}\right).
\end{eqnarray*}
Integration leads to
\begin{eqnarray}\nonumber
m&=&\frac{C^3}{6}{\int}A\left[3(Y_s+X_s)\frac{B'}{AB}
+6H(\Theta-\sigma_s)-3(Y_k-X_k)\frac{C'}{AC}\right.\\\nonumber
&+&{\kappa}q(2\sigma_s-\sigma_k+\Theta)+6H^*+3a(Y_s-Y_k-X_s+X_k)-\frac{{\kappa}s}{C^2}\\\label{72}
&\times&\frac{1}{{\pi}AB^2}\left(\left.s'-\frac{sC'}{C}
-\frac{4sB'}{B}\right)\right]dr+\frac{C^3\gamma(t)}{6},
\end{eqnarray}
where $\gamma$ is an arbitrary integration function of $t$. This
mass function shows its dependence on different factors, in
particular, on electric charge. Inserting the values of
$X_s,~Y_s,~X_k,~Y_k$ from Eq.(\ref{44}), it can be expressed in
terms of physical variables and the Weyl tensor
\begin{eqnarray}\nonumber
m&=&\frac{C^3}{2}\int A\left[\kappa(P_r-P_z)\frac{B'}{AB}
-\frac{2E_kC'}{AC}\right.+2
H(\Theta-\sigma_s)+\frac{{\kappa}q}{3}\\\nonumber
&\times&(2\sigma_s-\sigma_k+\Theta)+2a(E_s-E_k)\left.+2H^*\right]dr
-\frac{C^3}{6}\int\frac{{\kappa}s}{\pi B^2C^2}\\\label{73}
&\times&\left(s'-\frac{sC'}{C}\right.-\left.\frac{4sB'}{B}\right)dr+\frac{C^3\gamma(t)}{6}.
\end{eqnarray}
This gives the contribution of anisotropy, Weyl tensor, electric
charge and dissipation.

The transport equation for dissipative fluids is given by \cite{30}
\begin{equation}\label{74}
{\tau}h^{\alpha\beta}V^\gamma q_{\beta;\gamma}+q^\alpha
=-Kh^{\alpha\beta}(T_{,\beta}+Ta_\beta)
-\frac{1}{2}KT^2\left(\frac{{\tau}V^\beta}{KT^2}\right)_{;\beta}q^\alpha.
\end{equation}
Here $K,~T$ and $\tau$ indicate thermal conductivity, temperature
and relaxation time, respectively. The only one independent
component is
\begin{equation}\label{75}
{\tau}q^*+q=-K(T^\dag+Ta)-\frac{1}{2}KT^2q\left(\frac{\tau}{KT^2}\right)^*
-\frac{1}{2}q\tau\Theta.
\end{equation}
Substituting Eqs.(\ref{65}), (\ref{66}) and (\ref{75}) in
(\ref{57}), it follows that
\begin{eqnarray}\nonumber
&&(\mu+P_r)\left[1-\frac{KT}{\tau(\mu+P_r)}\right]U^*
=-(\mu+P_r)\left(\frac{{\kappa}P_rC^3}{2}+m\right)\frac{1}{C^2}\\\nonumber
&&\times\left[1-\frac{KT}{\tau(\mu+P_r)}\right]+\frac{C'}{A}\left[-P^\dag_r
+(P_z-P_r)\frac{B'}{AB}+(P_\phi-P_r)\frac{C'}{AC}\right]\nonumber
\end{eqnarray}
\begin{eqnarray}\nonumber
&&+\frac{C'}{A}\left[\frac{KT^2q}{2\tau}\left(\frac{\tau}{KT^2}\right)^*
+\frac{q}{\tau}+\frac{KT^\dag}{\tau}+\frac{1}{3}(\sigma_s
+\sigma_k)q-\left.\frac{5}{6}q\Theta\right]\right.\\\label{76}
&&+\frac{ss'}{\pi B^2C^2}
+\frac{{\kappa}s^2}{2{\pi}B^2C}(\mu+P_r)\left(1-\frac{KT}{\tau(\mu+P_r)}\right).
\end{eqnarray}
This shows that gravitational attraction and electric charge on any
fluid element will decrease by the same factor as the inertial mass
density.

We see that the charge increases the gravitational mass only if
\begin{equation}\label{77}
s'>s\kappa~C(\mu+P_r)\left(-1+\frac{KT}{\tau(\mu+P_r)}\right),
\end{equation}
otherwise, it will decrease. This increase of gravitational mass
causes rapid collapse. Bekenstein \cite{50} noticed this strange
effect for the Oppenheimer-Volkoff equations of hydrostatic
equilibrium. We see that the charge does not enter into the term
$1-\frac{K T}{\tau(\mu+P_r)}$ but it affects the gravitational mass
and shows how thermal effects reduce the effective inertial mass. We
observe that as $\frac{K T}{\tau(\mu+P_r)}\rightarrow1$, the
inertial mass density of the fluid element tends to zero \cite{36}.
This shows that there is no inertial force and matter would
experience a gravitational attraction which causes the collapse. For
$0<\frac{KT}{\tau(\mu+P_r)}<1$, the inertial mass density goes on
decreasing while $\frac{KT}{\tau(\mu+P_r)}>1$ indicates the increase
of inertial mass density. By the equivalence principle, there should
occur increase or decrease of mass. Thus one can easily distinguish
between expanding and collapsing mechanism during the dynamics of
dissipative process.

Assume that the collapsing cylinder evolves in such a way that the
value of $\frac{KT}{\tau(\mu+P_r)}$ increases and approaches to $1$
for some region. During this process, the gravitational force term
decreases and leads to a change of the sign of the right hand side
of Eq.(\ref{76}). This would happen for small values of the
effective inertial mass density and implies a strong bouncing of
that part of the cylinder \cite{37}. This phenomenon causes the loss
of energy from the system and hence the collapsing cylinder with
non-adiabatic source leads to the emission of the gravitational
radiations. Notice that the term $1-\frac{K T}{\tau(\mu+P_r)}$ (in
the inertial mass, gravitational force and electric charge) is
related to the left of Eq.(\ref{74}).

\section{Static Charged Anisotropic Cylinders}

Here we discuss all possible solutions of the field equations of the
static charged anisotropic cylinders. The corresponding field
equations are
\begin{eqnarray}\nonumber
{\kappa}A^2\left(\mu+\frac{s^2}{2{\pi}B^2C^2}\right)
&=&-\frac{B''}{B}-\frac{C''}{C}+ \frac{A'}{A}\left(\frac{B'}{B}
+\frac{C'}{C}\right)-\frac{B'}{B}\frac{C'}{C},\\\nonumber \kappa
A^2\left(P_r-\frac{s^2}{2{\pi}B^2C^2}\right)
&=&\frac{A'}{A}\left(\frac{B'}{B}+\frac{C'}{C}\right)+\frac{B'}{B}\frac{C'}{C},\\\nonumber
\kappa A^2\left(P_z+\frac{s^2}{2\pi B^2 C^2}\right)
&=&\frac{A''}{A}+\frac{C''}{C}-\left(\frac{A'}{A}\right)^2,\\\nonumber
\kappa A^2\left(P_\phi+\frac{s^2}{2{\pi}B^2C^2}\right)
&=&\frac{A''}{A}+\frac{B''}{B}-\left(\frac{A'}{A}\right)^2.\nonumber
\end{eqnarray}
These equations, in terms of auxiliary variables,
$\omega=\frac{A'}{A},~\xi=\frac{B'}{B},\quad\eta=\frac{C'}{C}$, can
be written as
\begin{eqnarray}\label{85}
{\kappa}A^2\left(\mu+\frac{s^2}{2{\pi}B^2C^2}\right)
&=&-\xi'-\xi^2-\eta'-\eta^2+\omega\xi+\omega\eta-\xi\eta,\\\label{86}
\kappa A^2\left(P_r-\frac{s^2}{2{\pi}B^2C^2}\right)
&=&\omega\xi+\omega\eta+\xi\eta,\\\label{87} \kappa
A^2\left(P_z+\frac{s^2}{2{\pi}B^2C^2}\right)
&=&\omega'+\eta'+\eta^2,\\\label{88}\kappa
A^2\left(P_\phi+\frac{s^2}{2{\pi}B^2C^2}\right)
&=&\omega'+\xi'+\xi^2.
\end{eqnarray}
Adding all these equations, we have
\begin{equation}
\omega'+\omega\xi+\omega\eta=Y_TA^2.\\\label{89}
\end{equation}
From Eqs.(\ref{86})-(\ref{88}), it follows that
\begin{eqnarray}\label{90}
\omega'+\xi'+\xi^2-\omega\xi-\omega\eta-\xi\eta
&=&\kappa(P_\phi-P_r)A^2+\frac{8s^2A^2}{B^2C^2},\\\label{91}
\omega'+\eta'+\eta^2-\omega\xi-\omega\eta-\xi\eta
&=&\kappa(P_z-P_r)A^2+\frac{8s^2A^2}{B^2C^2},\\\label{92}
\xi'+\xi^2-\eta'+\eta^2&=&\kappa(P_\phi-P_z).
\end{eqnarray}
In terms of auxiliary variables, the scalars $E_s$ and $E_k$ become
\begin{eqnarray}\nonumber
E_s&=&\frac{1}{2A^2}\left[-\omega'+\eta'+\eta^2+\omega\xi-\omega\eta-\xi\eta\right],\\\label{94}
E_k&=&\frac{1}{2A^2}\left[-\omega'+\xi'+\xi^2-\omega\xi+\omega\eta-\xi\eta\right].
\end{eqnarray}
Equations (\ref{44}), (\ref{90})-(\ref{94}) yield
\begin{equation}\label{95}
Y_sA^2=-\omega'+\omega\xi,\quad Y_kA^2=-\omega'+\omega\eta.
\end{equation}
Integration of the first equation yields
\begin{equation*}
A={\alpha}e^{{\int}B\left(\int\frac{-Y_sA^2}{B}dr\right)dr},
\end{equation*}
where $\alpha$ is an integration constant, which means $B=B(A)$ or
$\xi=\xi(\omega)$ for any $Y_s$. When we integrate second of the
above equation, we obtain
\begin{equation*}
A={\gamma}e^{{\int}C\left(\int\frac{-Y_kA^2}{C}dr\right)dr},
\end{equation*}
where $\gamma$ is another integration constant implying $C=C(A)$ or
$\eta=\eta(\omega)$ for any $Y_k$. Consequently we can express any
$(\omega,\xi,\eta)$ in terms of each other. This leads to the
similar result as that of paper \cite {26} with the effect of
charge. Thus any static anisotropic solution is determined by a
triplet of scalars $(Y_k,~Y_s,~X_k)$ or $(Y_k,~Y_s,~X_s)$ as in the
charged free case. Similarly, we can discuss the case of isotropic
cylinders.

\section{Conclusion}

This paper investigates the effects of electromagnetic field on
structure scalars of the cylindrically symmetric anisotropic
dissipative fluid. The electric charge increases the inhomogeneity
produced by local anisotropy. Following Herrera \cite{26}, we have
formulated dynamical as well as transport equations and also mass
function for the charged cylindrical system. It turns out that the
coupled dynamical-transport equation has an extra factor due to
charge. If we take $\frac{KT}{\tau(\mu+P_r)}=1$, the inertial mass
and gravitational force vanish while the gravitational mass reduces
for $s'<s\kappa C(\mu+P_r)\left(-1+\frac{KT}{\tau(\mu+P_r)}\right)$.
Further, we have discussed the static case in electromagnetic field
which shows that any solution of the field equations can be
expressed in terms of the scalar functions like charge free case. It
is worth mentioning here that all our results reduce to charge free
case \cite{26} for $s=0$.


\begin{thebibliography}{27}

\bibitem{1} Sawyer, R.F.: Phys. Rev. Lett. \textbf{29}(1972)382;
Letelier, P.: Phys. Rev. \textbf{D22}(1980)807; Herrera, L. and
Santos, N.O.: Phys. Rep. \textbf{286}(1997)53; Astrophys. J.
\textbf{438}(1995)308.

\bibitem{6} Lemaitre, G.: Ann. Soc. Sci. Bruxells \textbf{A53}(1933)51.

\bibitem{7} Bowers, R.L. and Liang, E.P.T.: Astrophys. J. \textbf{188}(1974)657.

\bibitem{8} Heintzmann, H. and Hillebrandt, W.: Astron. Astrophys.
\textbf{38}(1975)51; Stewart, B.W.: J. Phys. \textbf{A15}(1982)2419;
Bayin, S.: Phys. Rev. \textbf{D26}(1982)1262; Gokhroo, M.K. and
Mehra, A.L.: Gen. Relativ. Gravit. \textbf{26}(1994)75.

\bibitem{50} Bekenstein, J.D.: Phys. Rev. \textbf{D4}(1971)2185.

\bibitem{51} Oppenheimer, J.R. and Volkoff, G.: Phys. Rev. \textbf{55}(1939)374.

\bibitem{51a} Di Prisco, A., Herrera, L., Le Denmat, G., MacCallum, M. and Santos,
N.O.: Phys. Rev. \textbf{D76}(2007)064017;

\bibitem{52}  Nath, S., Debnath, U. and Chakraborty, S.: Astrophys.
Space Sci. \textbf{313}(2008)431.

\bibitem{53} Sharif, M. and Abbas, G.: Mod. Phys. Lett. \textbf{A24}(2009)2551;
J. Korean Physical Society \textbf{56}(2010)529; J. Phys. Soc. Jpn.
\textbf{80}(2011)104002; Sharif, M. and Siddiqa, A.: Gen. Relativ.
Gravit. \textbf{43}(2011)73;  Sharif, M. and Fatima, S.: Gen.
Relativ. Gravit. \textbf{43}(2011)127.

\bibitem{20} Bel, L.: Ann. Inst. H Poincar\'{e} \textbf{17}(1961)37.

\bibitem{21} Herrera, L., Ospino, J., Di
Prisco, A., Fuenmayor, E. and Troconis, O.: Phys. Rev.
\textbf{D79}(2009)064025; Herrera, L., Di Prisco, A., Ospino, J. and
Carot, J.: Phys. Rev. \textbf{D82}(2010)024021; Herrera, L., Di
Prisco, A. and Ospino, J.: Gen. Relativ. Gravit.
\textbf{42}(2010)1585.

\bibitem{25} Herrera, L., Di Prisco, A. and Ib\'{a}$\tilde{n}$ez, J.: Phys. Rev. \textbf{D84}(2011)107501.

\bibitem{26} Herrera, L., Di Prisco, A., and Ospino, J.: arXiv:1201.2862.

\bibitem{30} M\"{u}ller, I.: Z. Physik \textbf{198}(1967)329; Israel, W.: Ann. Phys.
\textbf{100}(1976)310; Israel, W. and Stewart, J.: Ann. Phys.
\textbf{118}(1979)341; Sharif, M. and Abbas, G.: Astrophys. Space
Sci. \textbf{335}(2011)515.

\bibitem{36} Herrera, L., Santos, N.O.: Phys. Rev. \textbf{D70}(2004)084004.

\bibitem{37} Herrera, L., Di Prisco, A and Barreto, W.: Phys. Rev. \textbf{D73}(2006)024008.

\end{thebibliography}
\end{document}